\documentclass[sigconf, nonacm, table,xcdraw, natbib=true]{acmart}

\usepackage{multirow}
\usepackage{enumitem}
\usepackage{balance}
\usepackage{soul}

\newcommand{\vect}[1]{\boldsymbol{#1}}

\newcommand{\matr}[1]{\boldsymbol{#1}}
\definecolor{DarkGreen}{HTML}{67eb2f}
\definecolor{LightGreen}{HTML}{97e077}

\AtBeginDocument{%
  }

\acmDOI{XXXXXXX.XXXXXXX}


\begin{abstract}

Cold-start challenges in recommender systems necessitate leveraging auxiliary features beyond user-item interactions. However, the presence of irrelevant or noisy features can degrade predictive performance, whereas an excessive number of features increases computational demands, leading to higher memory consumption and prolonged training times. 

To address this, we propose a feature selection strategy that prioritizes the user behavioral information. Our method enhances the feature representation by incorporating correlations from collaborative behavior data using a hybrid matrix factorization technique and then ranks features using a mechanism based on the maximum volume algorithm. This approach identifies the most influential features, striking a balance between recommendation accuracy and computational efficiency. We conduct an extensive evaluation across various datasets and hybrid recommendation models, demonstrating that our method excels in cold-start scenarios by selecting minimal yet highly effective feature subsets. Even under strict feature reduction, our approach surpasses existing feature selection techniques while maintaining superior efficiency.
\end{abstract}

\begin{document}

\title{Maximum Impact with Fewer Features: Efficient Feature Selection for Cold-Start Recommenders through Collaborative Importance Weighting}

\author{Nikita Sukhorukov}
\affiliation{%
  \institution{AIRI, Skoltech}
  \city{Moscow}
  \country{Russian Federation}
}
\email{sukhorukov@airi.net}

\author{Danil Gusak}
\affiliation{%
  \institution{AIRI, Skoltech}
  \city{Moscow}
  \country{Russian Federation}
}
\email{danil.gusak@skoltech.ru}

\author{Evgeny Frolov}
\affiliation{%
  \institution{AIRI, HSE University}
  \city{Moscow}
  \country{Russian Federation}
}
\email{frolov@airi.net}

\maketitle

\section{Introduction}

The task of generating relevant recommendations can be categorized into three distinct scenarios, each varying in the availability and nature of data. In the \emph{weak generalization scenario} (also known as standard scenario)~\cite{marlin_warm_start}, recommendations rely on a complete knowledge of past user-item interactions, focusing solely on known users or items present in the training data. In the \emph{strong generalization scenario} (or so-called warm-start)~\cite{marlin_warm_start}, one has to deal with new users or items that were not a part of the training data but are still associated with some known initial interactions. Finally, in the most extreme and challenging case of the \emph{cold-start scenario}~\cite{Aggarwal16, gusak2025dish}, newly introduced users or items lack any prior interaction history. In such cases, the prediction task requires extending the sources of knowledge about possible user preferences, since no direct interaction data are available \cite{Lam2008AddressingCP}. Understanding the effects of these sources of additional information on the quality of predictions becomes critical for successful accomplishment of the recommendation task.

These additional sources of knowledge are typically presented in the form of content features that describe users and items \cite{wang2021dcn, Shi2014CollaborativeFB}, and they are often referred to as metadata or side information \cite{Adomavicius2005TowardTN}.
Unlike the standard and warm-start scenarios, where the pure collaborative filtering approach suffices~\cite{gusak2024rece}, addressing cold start requires employing content-based or hybrid \cite{hybrid_rs_survey, bernardis2018novel} models capable of leveraging side features for accurate predictions in the absence of interaction data. 
However, empirical studies consistently demonstrate that pure collaborative filtering approaches outperform content-based models \cite{glauber2019collaborative, Lops2011ContentbasedRS} when such comparison is possible, i.e. when the collaborative signal is fully available but is not necessarily used, depending on the type of recommendation algorithm~\cite{evenafewratingsmorevaluable}.

For instance, a large feature space can make it challenging to differentiate relevant signals from noise, while also increasing computational complexity \cite{Blum1997SelectionOR, Guyon2003AnIT}.
Exceedingly large feature spaces raise the risks of overfitting to uninformative signals, resulting in lower generalization capabilities of recommendation algorithms~\cite{li_causal_fm, Rendle2012FactorizationMW}. Even in the absence of noise, side features may not fully represent factors affecting user choices in collaborative behavior and sometimes may \emph{even contradict it}~\cite{wang_causal_inference}. This results in an \emph{overspecialization effect}, which is particularly pronounced in purely content-based approaches, thereby hindering the predictive capabilities of models that heavily depend on side information.

These observations not only underscore the importance of careful consideration and curation of features, but also highlight an important challenge. The signal derived from \emph{content features may not align well with patterns observed in user-item interactions}. As the latter represents the underlying true user preferences, a good selection algorithm must ensure alignment of the selected features with the collaborative signal. Failure to filter unimportant features and prioritize the most informative ones is likely to degrade recommendations quality~\cite{wegmeth2022impact}.
This task is particularly vital in the cold-start regime, where recommendations rely solely on side information and are highly sensitive to undesirable deviations in the input data. 

Recognizing this challenge has guided recent research towards a particular problem formulation aimed at ensuring that the correlations induced by side information are compatible with those manifested by behavioral data~\cite{Nembrini_CQFS}. In this work, we adopt the same core principle but examine it from a novel angle. For this purpose, we propose a two-step feature selection approach. 
In the first step, we reframe the use of a hybrid recommender model called HybridSVD~\cite{HybridSVD, ALMEIDA2022116335, 10.1145/3404835.3462958}, enabling the \emph{propagation of correlations coming from collaborative signal into the latent representations of side features}. 

In the second step, we rerank the resulting feature embeddings using the rectangular Maximal-Volume (MaxVol) algorithm~\cite{mikhalev2018rectangular, goreinov2010find}. It is designed to \emph{utilize the enriched latent information obtained from the previous step to promote the most representative features} in terms of their consistency with user behavior.

This two-step process allows identifying the essential features that potentially improve recommendation quality. Moreover, limiting the number of involved features with proper selection methods can significantly reduce the training times of the recommender models \cite{Blum1997SelectionOR, Guyon2003AnIT}. In summary, our main contributions are:
\begin{itemize}
[leftmargin=*]
    \item We introduce a novel two-step feature selection approach that identifies the subset of features most aligned with user behavior;
    \item We evaluate our approach under extreme conditions where the number of available features is limited;
    \item We demonstrate that our approach surpasses other feature selection algorithms in terms of recommendation quality, requiring significantly fewer features.

\end{itemize}

\section{Collaborative Data-Driven Feature Selection}
\label{sec:feature_selection}

It is commonly observed that even a few ratings can be more important in terms of recommendation accuracy than side information \cite{evenafewratingsmorevaluable}. Collaborative filtering leverages the ``wisdom of crowds'' phenomenon \cite{mavrodiev2012effects} to recognize individual patterns and similarities in collective behavior, thus yielding more reliable recommendations \emph{closely aligned with underlying users' decision-making processes}. In contrast, side information in the form of features is often \emph{incomplete and noisy}~\cite{martinezt_noisy_incomplete_data, Lops2011ContentbasedRS}, providing only broad or stereotypical knowledge that does not always capture the diverse range of user preferences. Therefore, it is reasonable to \emph{prioritize collaborative signals over side information}, ensuring that any additional feature sources complement rather than contradict the observed user behaviors. This also prescribes the recipe for feature selection: the best candidates for the selection are the features that demonstrate the strongest correlation with user-item interactions.

\begin{figure}
\setlength{\abovecaptionskip}{0.1pt}
\centering
  \includegraphics[width=0.75\columnwidth]{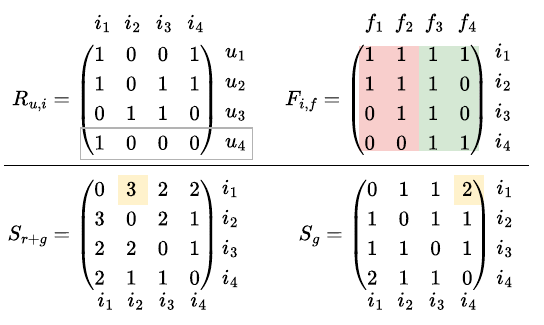}
  \caption{Example illustrating the effect of feature selection on item similarity. Item similarity matrices built using all features ($S_{r+g}$) and only green features ($S_g$). Yellow highlights the item most similar to item $i_1$, which was previously consumed by the user $u_4$.}
  \label{fig:example}
\end{figure}

To illustrate the challenge of feature selection in recommender systems, consider the toy example in Fig.~\ref{fig:example}, which shows a user-item interaction matrix $R_{u,i}$ and an item-feature matrix $F_{i,f}$. We aim to generate recommendations for user $u_4$ using a simple content-based KNN approach \cite{itemknn}, where item similarity is computed based on their feature vectors ($\matr{S}=\matr{F}_{i,f}\matr{F}_{i,f}^\top$). If we use the full feature set, the resulting item scores for user $u_4$ are $[0, 3, 2, 2]$. In this case, the model recommends item $i_2$.
However, if we restrict the model to use only features $f_3$ and $f_4$ (green columns), excluding the rest, the score distribution changes to $[0, 1, 1, 2]$, and the recommended item becomes $i_4$. Looking back at the interaction matrix $R_{u,i}$, we see that items $i_1$ and $i_2$ are never co-consumed, while $i_1$ and $i_4$ are. This suggests that the reduced feature set (green features) yields recommendations more aligned with actual user behavior.

Motivated by these insights, we aim to design a new feature selection method that explicitly incorporates the described notion of compatibility into the selection process in a straightforward way. Although prior work has explored the integration of side information into recommender models, \emph{feature selection itself remains an open problem, particularly in cold-start scenario}. 

In this paper, we adapt the robust and versatile HybridSVD \cite{HybridSVD, ALMEIDA2022116335} approach to accomplish this task. In the original work, the authors employ a special generalized SVD-based formulation to incorporate additional feature-based correlations into the latent representations of users and items to mitigate sparsity and address cold start problems. However, our use of HybridSVD deviates significantly from its original purpose and application. Unlike the standard top-$N$ recommendation setup, \emph{our objective is not to improve item ranking, but to identify which features are most aligned with collaborative behavior}. Our approach ``inverts'' the usual HybridSVD workflow by using the feature space as the primary focus and enriching it with the collaborative correlations derived from user-item interactions. This procedure allows the resulting latent representations of features to be more \emph{aligned with the observed collaborative behavior}.

\begin{figure*}
\setlength{\abovecaptionskip}{4pt}
\centering
  \includegraphics[width=0.85\textwidth]{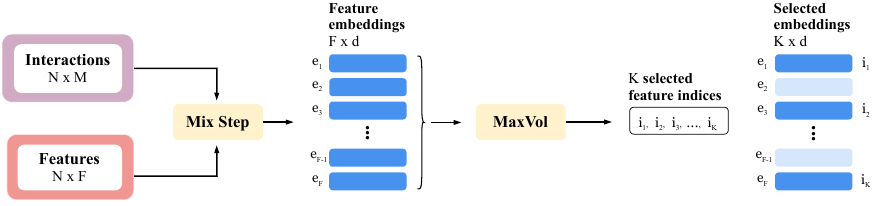}
  \caption{Proposed feature selection approach. The collaborative information is embedded into the features through HybridSVD procedure at the Mix step (see Section~\ref{hybridsvd selection}). The resulting collaboratively enriched feature embeddings are passed to the MaxVol algorithm that selects the most representative set of features. The selected feature embeddings are highlighted in blue, corresponding to the relevant columns in the Features matrix.}
  \label{fig:feature_selection}
\end{figure*}

Once the features are enriched with behavioral information in their latent space and are aligned with the signal from the collaborative component, we rank them based on their importance. To formalize the notion of ``importance'' in this new latent representation, we employ a natural algebraic approach. We propose selecting features whose latent vectors \emph{form the most representative basis, spanning the majority of the latent space}. This is achieved by the means of the rectangular MaxVol \cite{teneva1, mikhalev2018rectangular, goreinov2010find} algorithm, which is widely used in the field of recommender systems for tasks such as rating elicitation~\cite{liu_maxvol_2011, fonarev_maxvol_2016}. By construction, such basis contains the most descriptive elements that maximize knowledge about the feature space, enriched with the collaborative correlations through the HybridSVD scheme. Thus, we posit that the selected basis captures user preferences and provides the necessary feature compatibility. The visual representation can be found in Figure~\ref{fig:feature_selection}.

\subsection{Integration of Collaborative Data} \label{hybridsvd selection}

Let us denote $N$ as the number of items, $M$ as the number of users, and $F$ as the number of features; the item-feature matrix is denoted as $\matr{F}\in\mathbb{R}^{N\times F}$, where each row represents a feature vector. First, we apply TF-IDF~\cite{manning2008introduction} weighting to the item-feature matrix: 
\begin{equation}
\label{eq:tfidf}
    \matr{F}_\text{w} = \text{TF-IDF}(\matr{F})
\end{equation}
This reweighting reduces the impact of frequently occurring features and highlights those that are relatively rare but potentially more descriptive and informative from an information retrieval perspective \cite{manning2008introduction}.
By emphasizing these unique features, TF-IDF weighting helps to capture the most relevant aspects of the data, thereby enhancing the effectiveness of our feature selection process.
Moreover, such weighting aligns with collaborative behavior, as features that uniquely characterize items are more likely to correlate with user preferences reflected in interaction data.

We construct the collaborative similarity matrix $\matr{S}\in\mathbb{R}^{N\times N}$ that measures the similarity between items based on collaborative data, specifically user-item interactions:
\begin{equation}
    \matr{S} = (1 - \alpha) \matr{I} + \alpha \text{Sim}(\matr{R}),
\end{equation}
where $\matr{R}\in\mathbb{R}^{M\times N}$ is the sparse user-item interactions matrix; $\matr{I}$ is the identity matrix; the $\text{Sim}(\cdot)$ function computes similarity scores between each pair of items; and $\alpha \in [0, 1]$ is a hyperparameter that controls an influence of the collaborative part via off-diagonal elements. When $\alpha$ is set to 0, we do not use collaborative information at all, and the approach will rely solely on side information. As $\alpha$ increases, the influence of the behavioral signals from users grows, thereby incorporating more of the collaborative filtering component into the similarity scores. This allows the similarity matrix to blend both side information and collaborative signals, adjusting the balance according to the value of $\alpha$.

In our experiments, we compute collaborative item similarity using the cosine similarity function applied to the columns of the interaction matrix $\matr{R}$. Other symmetric similarity functions could also be employed to measure item similarity. Also, utilizing pretrained collaborative filtering models and their item collaborative similarity matrix can potentially enhance similarity measures. Exploring these alternatives is left for future investigation.

\subsubsection{Injection of collaborative signal}

We enhance the feature embeddings by propagating the collaborative signal through the Cholesky factor $\matr{L}$, while also incorporating item popularity information, which is often a strong predictor of user-item interactions. This is achieved via the following transformation:
\begin{equation}
    \matr{F}_{\text{sat}} = \matr{D}^p\matr{L}^\top\matr{F}_{\text{w}}, \ \ \ \ \ \matr{S}=\matr{L}\matr{L}^\top,
\end{equation}
where $\matr{L} \in \mathbb{R}^{N\times N}$ is a lower triangular matrix obtained from the Cholesky decomposition~\cite{golub2013matrix} of item collaborative similarity matrix $\matr{S}$; $\matr{D}\in \mathbb{R}^{N\times N}$ is a diagonal matrix containing the number of interactions with each item, $p$ serves as a hyperparameter. When $p>0$, features of popular items are amplified, promoting their importance in the embedding space. When $p<0$, greater weight is assigned to features of less popular (long-tail) items.

\subsubsection{Feature embedding derivation.}
To get the feature embeddings we perform the truncated SVD~\cite{golub2013matrix} of the matrix $\matr{F}_{\text{sat}}$ of rank $k$, which is also a hyperparameter:
\begin{equation}
    \matr{F}_{\text{sat}} = \matr{U} \matr{\Sigma} \matr{V}^\top,
\end{equation}
where $\matr{U} \in \mathbb{R}^{N \times k}, \matr{\Sigma}\in \mathbb{R}^{k \times k}, \matr{V} \in \mathbb{R}^{F \times k}$. 

Finding the low-rank decomposition of the $\hat{\matr{F}}$ achieves two goals:
\begin{itemize}[leftmargin=*]
    \item SVD allows adjusting the amount of variability in data that we capture, keeping the most critical information while discarding less important variations corresponding to the smallest singular values~\cite{ekstrand}. By selecting an appropriate $k$, we can ensure that the essential structure of the data is preserved while reducing noise and redundancy.
    \item Most recommendation datasets contain more items than features, so we ensure that the feature embeddings matrix $\matr{V}\in\mathbb{R}^{F\times k}$ will be tall ($F > k$). We need this condition for the next step of our feature selection approach (see Section~\ref{maxvol selection}).
\end{itemize}

In total, our approach includes three hyperparameters: $\alpha$ -- contribution of collaborative signal, $p$ -- contribution of item popularity, and $k$ -- dimensionality of feature embeddings. The transformations described in this section are collectively referred to as the Mix step, as illustrated in Figure~\ref{fig:feature_selection}. 

\subsection{MaxVol-based Feature Selection} \label{maxvol selection}

The transformations described in the previous section allow us to derive efficient low-dimensional feature embeddings $\matr{V}$. Matrix $\matr{V}$ captures both the intrinsic characteristics of items and latent patterns of interactions, which are critical for addressing the cold-start problem. The next step is to select a subset that is most beneficial for top-n recommendation task. One of the most theoretically grounded approaches for this task is the MaxVol algorithm \cite{goreinov2010find, mikhalev2018rectangular, mezentsev2024scalable}. 

The algorithm finds a submatrix of a maximal volume in a tall matrix, which is the case for $\matr{V}$ being of the shape $F \times k$, $F > k$. The volume of the matrix $\matr{V}$ is defined through a determinant as $\text{Vol}(\matr{V})=\sqrt{|\matr{V}^\top\matr{V}|}$.
This algorithm selects a subset of vectors that maximizes the volume of the parallelepiped formed by the chosen vectors in the embedding space. By doing so, it ensures that the selected features are diverse and representative as a set, capturing the most informative dimensions of the data.

We note that the standard (square) MaxVol algorithm can only find a square $k \times k$ submatrix of maximum volume, which is not suitable for the selection of an arbitrary number of features. To alleviate this shortcoming, we employ the \emph{rectangular} MaxVol\footnote{We used implementation from \url{https://github.com/AndreiChertkov/teneva}}, which finds a rectangular submatrix. The matrix has to be tall, i.e., have more rows than columns, otherwise, the volume of such matrix will be 0. The algorithm starts with a non-singular submatrix $\hat{\matr{V}}$ of size $k \times k$ and iteratively swaps the rows with the rest of a matrix to maximize $\text{Vol}(\hat{\matr{V}})$, where the accuracy is controlled by a preset parameter. After the square $k \times k$ submatrix is found, it proceeds to greedily append leftover rows from $\matr{V}$ to $\hat{\matr{V}}$ until the required number of rows (features) is reached. 

\begin{figure}[tb!]
\setlength{\abovecaptionskip}{7pt}
   \centering
   \includegraphics[width=0.52\columnwidth]{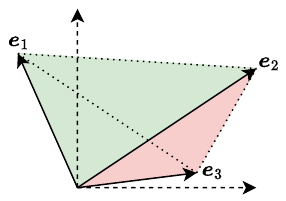}
   \Description{A figure describing the MaxVol algorithm.}
   \caption{Intuition behind MaxVol algorithm. The triangle formed by vectors $\boldsymbol{e}_1$ and $\boldsymbol{e}_2$ has the largest area among all possible pairs, indicating that these vectors span the most representative subspace and capture the most information.
   }
   \label{fig:maxvol}
 \end{figure}

A visual representation of the MaxVol algorithm’s underlying principles is shown in Figure \ref{fig:maxvol}. In our approach, feature embeddings are augmented with collaborative signals derived from interaction data. The MaxVol algorithm is then applied in this modified latent space to identify the most informative features.

As a result, the algorithm outputs a ranked list of features, reflecting how much each row (feature embedding) of $\matr V$ contributes to preserving the variance. In other words, selecting high-volume submatrices identifies the features that best span the feature embedding space, thereby defining their importance. We call this \emph{collaborative feature weighting} due to the injection of the collaborative information into the feature space. 

\section{Experiments}
\label{sec:expers}

In this section, we describe the experiments that evaluate the performance of various feature selection algorithms and conduct a comprehensive comparison, following the setup proposed in \cite{Nembrini_CQFS}.

\subsection{Experimental Settings}
\label{sec:expsetup}

\subsubsection{Datasets}
We evaluate our feature selection algorithm and baselines on seven widely used datasets: \emph{MovieLens-1M} (ML-1M)~\cite{ml1m} and \emph{Book Crossing} (BX)~\cite{bookcrossing}, which contain user ratings for movies and books, respectively; \emph{MTS Library} (MTS)~\cite{mts_library}, featuring user–book interactions and rich metadata; \emph{CiteULike-a} (CiteUL)~\cite{cite_dataset}, a scientific paper dataset with tag-based features from titles and abstracts; and three Amazon domains: \emph{Video Games} (Amz G), \emph{Industrial Scientific} (Amz IS), and \emph{Software} (Amz S)~\cite{amazon_reviews}, which include user reviews and semantically rich textual features.

These datasets span diverse domains, interaction types, and feature modalities, enabling a comprehensive evaluation of the effectiveness and generalizability of our feature selection approach. Table~\ref{tab:datasets} summarizes their key characteristics.

\begin{table}[b]
\setlength{\abovecaptionskip}{5pt}
  \caption{General information about datasets. ``nnz'' is density of known elements in the data.}
  \label{tab:datasets}
  \resizebox{\columnwidth}{!}{%
  \begin{tabular}{lccccl}
    \toprule
    Dataset & Users & Items & nnz, \% & Features & Categories\\
    \midrule
    ML-1M~\cite{ml1m} & 6,040 & 3,706 & 4.50 & 3,665 & Actors, writers, directors\\
    BX~\cite{bookcrossing} & 9,966 & 14,368 & 0.10 & 2,385 & Authors, publishers\\
    MTS~\cite{mts_library} & 7,916 & 9,318 & 0.15 & 1,271 & Genres, authors\\
    \midrule
    CiteUL~\cite{cite_dataset} & 16,980 & 5,551 & 0.22 & 8,000 & Tags\\
    Amz G \cite{amazon_reviews} & 12,805 & 5,680 & 0.30 & 9,289 & Reviews\\
    Amz IS \cite{amazon_reviews} & 54,353 & 27,199 & 0.03 & 7,661 & Reviews\\
    Amz S \cite{amazon_reviews} & 30,785 & 6,462 & 0.25 & 5,448 & Reviews\\
    \bottomrule
  \end{tabular}%
  }
\end{table}

\subsubsection{Feature Preprocessing}
\label{datasets-feature_preprocessing}

For datasets with categorical (non-textual) features (ML-1M, MTS, BX) we apply one-hot encoding and remove features appearing in fewer than two items, retaining only those with meaningful co-occurrence patterns. In the Amazon datasets, we treat item reviews as textual features. Reviews are concatenated per item, cleaned, tokenized, stemmed, and transformed via count vectorization. Tokens occurring fewer than 10 times are discarded. This preprocessing yields significantly larger feature spaces compared to the non-textual datasets.

\begin{figure}
\setlength{\abovecaptionskip}{7pt}
   \centering
   \includegraphics[width=\columnwidth]{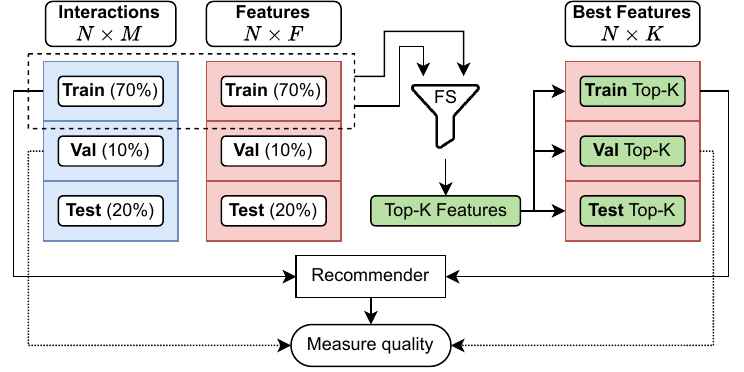}
   \Description{A figure describing the process of the recommender model optimization.}
   \caption{The pipeline for finding the best model on the validation set. The interaction data is split by items. Train interactions and train item features are used to determine the set of the most important features, using the chosen feature selection approach depicted as funnel labeled FS. The recommender model is trained using the train interactions and train item features, the quality is evaluated using interactions and selected item features from validation set.}
   \label{fig:pipeline}
 \end{figure}

\subsection{Data Splitting}
\label{sec:data_splitting}

We split the data based on items into train, validation, and test sets using a $(70\% / 10\% / 20\%)$ proportion, respectively. Cold-start items for the validation and test sets are randomly sampled, while the remaining items and their interactions form the train set.  As a result, we get three interaction sets disjoint by items, ensuring that the feature matrices corresponding to the training, validation, and test items are also disjoint. Since we address a pure cold-start scenario, cold items are described solely by their features. This strict setup allows us to isolate the effect of injecting collaborative information into feature selection, avoiding interference from warm-item interactions that could obscure the true impact of our method.

\subsubsection{Evaluation}
\label{sec:eval}

We use Recall@$10$ (R@$10$) as the target metric for model optimization, calculating it against all interactions in the holdout set~\cite{timetosplit2025, 2024autoregressive}. The results for MRR and Coverage metrics, along with code for reproducing our work, will be soon published on GitHub.

\subsubsection{Implementation Details}
\label{sec:implementation_details}

The experiment pipeline is depicted in Figure~\ref{fig:pipeline}. We perform random search~\cite{bergstra_random_search} over 20 sampled configurations, drawn from the Cartesian product of model and feature selection hyperparameter grids. This joint tuning strategy enables simultaneous optimization of both model performance and feature selection quality. To ensure statistical robustness, all experiments are repeated across 10 random data splits.

\subsubsection{Recommender Models}
\label{sec:recommenders}

To measure the performance of our feature selection approach and other baseline algorithms, we use several hybrid recommender models, as they typically show better quality compared to pure content-based approaches~\cite{adomavicius_tuzhilin_toward_next_gen_rec_sys}. However, we included a content-based model in our experiments because of its simplicity and as a sanity check for other hybrid models.

The models we use in our experiments: \emph{LightFM}~\cite{LightFM}, a model that is a special case of Factorization Machines~\cite{FM_rendle}; \emph{Local Collective Embeddings (LCE)}~\cite{LCE}, designed to capture local interaction patterns; \emph{HybridSVD}~\cite{HybridSVD}, a generalization of PureSVD~\cite{cremonesi_puresvd} that incorporates side information; and \emph{ItemKNN}~\cite{itemknn}, using a content-based variant with cosine similarity over feature vectors. We focus on classical models because they tend to perform better in cold-start scenarios, where data is sparse and noisy, and where deep learning approaches often struggle due to overfitting or insufficient signal~\cite{puthiya2020simple,ferrari2019we}.

\subsubsection{Feature Selection Baselines}
\label{sec:selection_baselines}

We compare the performance of our approach to several baseline algorithms:

\begin{itemize}[leftmargin=*]
    \item \textbf{All features.} This approach uses all available features to fully assess the performance of the model without feature selection. This helps us to understand trade-offs between feature selection and model performance, showing how the number of features affects accuracy and efficiency of a recommender system.

    \item \textbf{Random.} This baseline selects a fixed number of features uniformly at random from the full feature set. It serves as a control to benchmark the effectiveness of systematic feature selection methods. Comparing model performance with randomly chosen features helps quantify the value added by more informed selection strategies.

    \item \textbf{Collaborative feature popularity (Popular).} This baseline assumes that user interactions with items implicitly signal interactions with their associated features. Thus, features linked to more popular items are considered more important. We compute feature importances as $\vect{w} = \vect{e}^\top\matr{R}\matr{F}$.
    where $\vect{e}^\top\matr{R}$ yields item popularity (interaction counts), and the product with $\matr{F}$ aggregates these counts over features. The resulting vector is sorted in non-increasing order to get the indices of the most important features. It offers a straightforward yet effective approach to feature selection by incorporating collaborative user behavior.

    \item \textbf{CFeCBF}. This approach is based on the idea of compatibility between collaborative and content information \cite{Deldjoo_cfecbf}. The weighting of features is learned by approximating the collaborative item similarity matrix $\matr{S}^{\text{CF}}$ with the content-based item similarity matrix $\matr{S}^{\text{CBF}}$ as follows:
    \begin{equation}
    \label{eq:cfecbf}
    \arg\min_{\vect{w}}{\|\matr{S}^{\text{CF}} - \matr{S}^{\text{CBF}}\|_F + \lambda_1 \|\vect{w}\|_1 + \lambda_2 \|\vect{w}\|_2^2}
    \end{equation}
    with $\matr{S}^{\text{CBF}}$ computed from normalized item-feature matrix $\widetilde{\matr{F}} = \text{diag}(\matr{F}\matr{F}^\top)^{-\frac{1}{2}}\matr{F}$ as $\matr{S}^{\text{CBF}} = \widetilde{\matr{F}}\,\text{diag}({\vect{w}})\,\widetilde{\matr{F}}^\top$. Collaborative similarity matrix $\matr{S}^{\text{CF}}$ is constructed using collaborative filtering recommender model. We use the learned weights to rank the features.

\end{itemize}

All features, random and popular baselines operate on the original feature matrix $\matr{F}$. Random and collaborative feature popularity baselines have not been extensively studied in the literature. Despite their simplicity and ease of implementation, they serve as valuable benchmarks for evaluating more complex feature selection approaches. Their straightforward nature makes them ideal for establishing reference points and assessing the relative performance of advanced methods in feature selection.
\subsection{Results}
\label{sec:results}

\subsubsection{Comparison with Other Feature Selection Approaches}

\begin{figure}[]
\setlength{\abovecaptionskip}{4pt}
   \centering
   \includegraphics[width=\columnwidth]{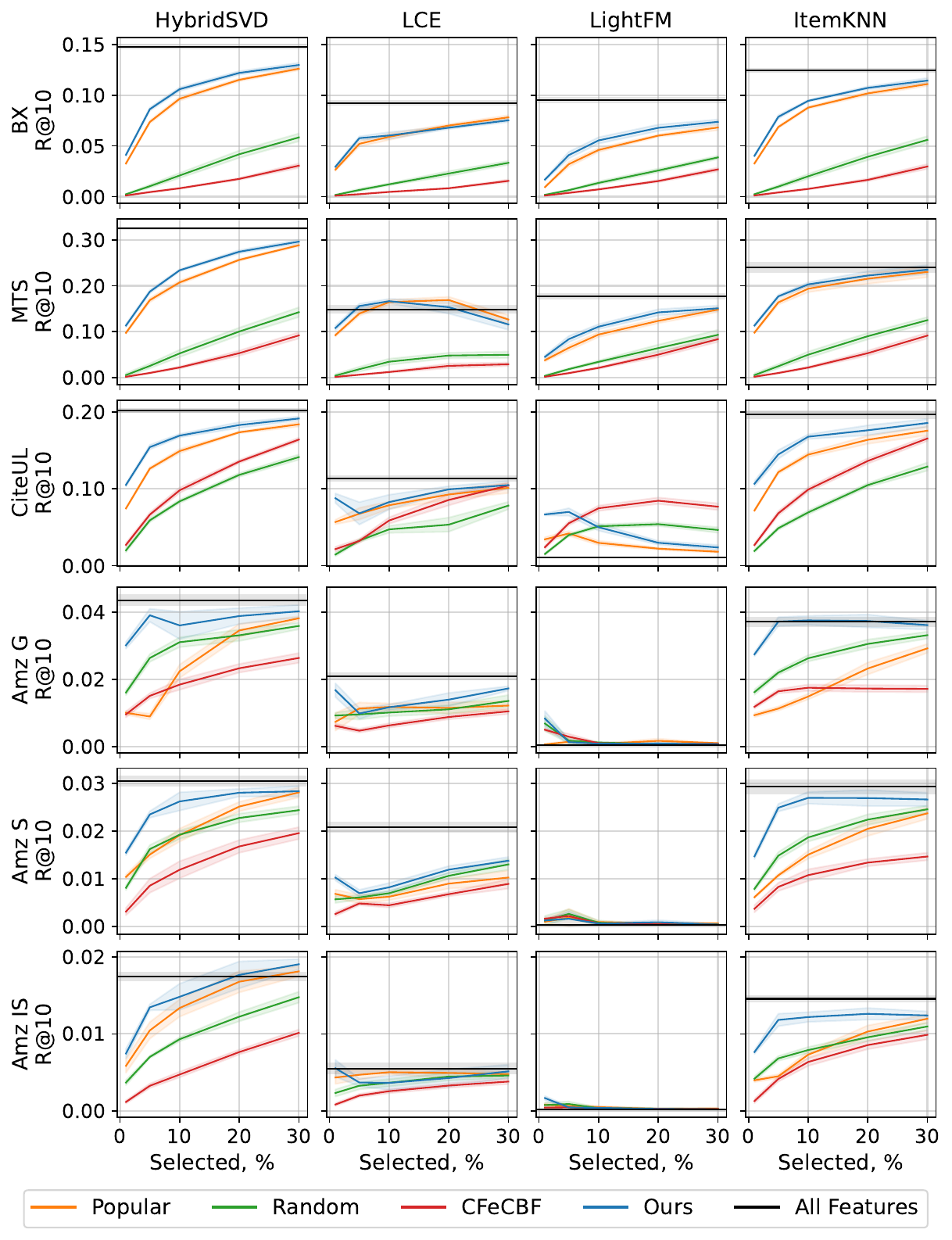}
   \Description{A figure with performance comparison of different feature selection algorithms.}
   \caption{The performance of different feature selection algorithms with respect to Recall@10.}
   \label{fig:results_recall}
 \end{figure}

In the plots of this section, the x-axis represents the percentage of features selected, and the y-axis shows the measured metric. A black horizontal line indicates performance using all features, while the shaded areas represent the 95\% confidence interval. Figure~\ref{fig:results_recall} presents the results in terms of Recall@10, with a focus on selecting up to 30\% of features to explore extreme cases. The HybridSVD and ItemKNN models generally showed better quality than the LCE and LightFM models, especially on datasets with textual features. Notably, LightFM model always exhibited significant overfitting on datasets with textual features (Amazon and CiteULike datasets), where the number of features is large, but exploring the reasons behind such behavior is beyond the scope of this work.

Table~\ref{tab:results} summarizes the results, showing the relative performance improvement (in percent) of our approach compared to the best-performing feature selection baseline (random, popular, or CFeCBF). Performance is evaluated using Recall@10. Note that the values reflect relative improvements, not the absolute metric scores. LightFM was excluded from Amazon datasets due to overfitting. It is evident that in most cases, our approach shows statistically significant improvement over the baselines. Even in cases where our approach falls behind at certain feature percentages (Amz G and Amz IS), Figure~\ref{fig:results_recall} illustrates that LCE model with our feature selection approach at 1\% of features achieves quality either better or comparable to other baselines across all feature percentages. A similar trend is observed with the MTS dataset, where our approach outperformed LCE model trained on all available features at 10\% of selected features and matched the quality of the popular baseline at 10\% and 20\%. LightFM model on the CiteULike dataset exhibits significant overfitting: our approach demonstrates a 533\% improvement over the model trained on all available features (Fig.~\ref{fig:results_recall}) and almost 2 times improvement over the best-performing popular baseline (Table~\ref{tab:results}), \emph{using just 1\% of features}. In contrast, the CFeCBF baseline requires 10\% of features to reach this level of performance. These results underscore that our approach performs best at low feature percentages, where precise feature selection is crucial.

\begin{table}
\setlength{\abovecaptionskip}{5pt}
\caption{Relative improvement (in \%) of the proposed method over the best-performing feature selection baseline (Random/Popular/CFeCBF) in terms of Recall@10. \textcolor{DarkGreen}{Deep green} indicates statistically significant improvements ($p\text{-value}<0.05$), \textcolor{LightGreen}{faded green} denotes insignificant improvements.}
\label{tab:results}
\scriptsize
\resizebox{\columnwidth}{!}{%
\begin{tabular}{llrrrrr}
\hline
 &
   &
  \multicolumn{5}{c}{Selected Features, \%} \\
\multirow{-2}{*}{Dataset} &
  \multirow{-2}{*}{Model} &
  1\% &
  5\% &
  10\% &
  20\% &
  30\% \\ \hline
 &
  HybridSVD &
  \cellcolor[HTML]{D1F5C7}17.27 &
  \cellcolor[HTML]{D1F5C7}18.43 &
  \cellcolor[HTML]{D1F5C7}21.75 &
  \cellcolor[HTML]{D1F5C7}10.59 &
  \cellcolor[HTML]{D1F5C7}15.82 \\
 &
  LCE &
  \cellcolor[HTML]{D1F5C7}16.04 &
  \cellcolor[HTML]{EBF6E8}11.68 &
  \cellcolor[HTML]{EBF6E8}6.03 &
  \cellcolor[HTML]{EBF6E8}6.05 &
  \cellcolor[HTML]{EBF6E8}6.21 \\
 &
  LightFM &
  \cellcolor[HTML]{D1F5C7}16.86 &
  \cellcolor[HTML]{D1F5C7}18.67 &
  \cellcolor[HTML]{D1F5C7}16.32 &
  \cellcolor[HTML]{D1F5C7}13.94 &
  \cellcolor[HTML]{D1F5C7}14.09 \\
\multirow{-4}{*}{ML-1M} &
  ItemKNN &
  \cellcolor[HTML]{D1F5C7}22.21 &
  \cellcolor[HTML]{D1F5C7}31.79 &
  \cellcolor[HTML]{D1F5C7}30.34 &
  \cellcolor[HTML]{D1F5C7}23.82 &
  \cellcolor[HTML]{D1F5C7}15.73 \\ \hline
 &
  HybridSVD &
  \cellcolor[HTML]{D1F5C7}25.98 &
  \cellcolor[HTML]{D1F5C7}17.11 &
  \cellcolor[HTML]{D1F5C7}9.86 &
  \cellcolor[HTML]{D1F5C7}5.95 &
  \cellcolor[HTML]{D1F5C7}2.92 \\
 &
  LCE &
  \cellcolor[HTML]{D1F5C7}10.94 &
  \cellcolor[HTML]{D1F5C7}10.73 &
  \cellcolor[HTML]{EBF6E8}2.27 &
  \cellcolor[HTML]{EBF6E8}-3.05 &
  \cellcolor[HTML]{EBF6E8}-3.89 \\
 &
  LightFM &
  \cellcolor[HTML]{D1F5C7}78.54 &
  \cellcolor[HTML]{D1F5C7}28.06 &
  \cellcolor[HTML]{D1F5C7}20.57 &
  \cellcolor[HTML]{D1F5C7}12.88 &
  \cellcolor[HTML]{D1F5C7}8.21 \\
\multirow{-4}{*}{BX} &
  ItemKNN &
  \cellcolor[HTML]{D1F5C7}22.83 &
  \cellcolor[HTML]{D1F5C7}14.75 &
  \cellcolor[HTML]{D1F5C7}7.62 &
  \cellcolor[HTML]{D1F5C7}5.49 &
  \cellcolor[HTML]{D1F5C7}3.17 \\ \hline
 &
  HybridSVD &
  \cellcolor[HTML]{D1F5C7}16.35 &
  \cellcolor[HTML]{D1F5C7}11.02 &
  \cellcolor[HTML]{D1F5C7}12.81 &
  \cellcolor[HTML]{D1F5C7}6.94 &
  \cellcolor[HTML]{D1F5C7}2.78 \\
 &
  LCE &
  \cellcolor[HTML]{D1F5C7}16.99 &
  \cellcolor[HTML]{D1F5C7}11.86 &
  \cellcolor[HTML]{EBF6E8}1.27 &
  \cellcolor[HTML]{EBF6E8}-9.32 &
  \cellcolor[HTML]{EBF6E8}{\ul -8.54} \\
 &
  LightFM &
  \cellcolor[HTML]{D1F5C7}19.57 &
  \cellcolor[HTML]{D1F5C7}29.23 &
  \cellcolor[HTML]{D1F5C7}18.34 &
  \cellcolor[HTML]{D1F5C7}15.08 &
  \cellcolor[HTML]{D1F5C7}\textit{1.96} \\
\multirow{-4}{*}{MTS} &
  ItemKNN &
  \cellcolor[HTML]{D1F5C7}15.72 &
  \cellcolor[HTML]{D1F5C7}8.09 &
  \cellcolor[HTML]{D1F5C7}4.91 &
  \cellcolor[HTML]{EBF6E8}{\ul 3.25} &
  \cellcolor[HTML]{EBF6E8}{\ul 2.23} \\ \hline
 &
  HybridSVD &
  \cellcolor[HTML]{D1F5C7}41.06 &
  \cellcolor[HTML]{D1F5C7}22.19 &
  \cellcolor[HTML]{D1F5C7}13.52 &
  \cellcolor[HTML]{D1F5C7}5.47 &
  \cellcolor[HTML]{D1F5C7}4.18 \\
 &
  LCE &
  \cellcolor[HTML]{D1F5C7}54.31 &
  \cellcolor[HTML]{EBF6E8}0.74 &
  \cellcolor[HTML]{EBF6E8}5.29 &
  \cellcolor[HTML]{EBF6E8}7.26 &
  \cellcolor[HTML]{EBF6E8}3.62 \\
 &
  LightFM &
  \cellcolor[HTML]{D1F5C7}94.43 &
  \cellcolor[HTML]{D1F5C7}26.55 &
  \cellcolor[HTML]{D1F5C7}-33.17 &
  \cellcolor[HTML]{D1F5C7}-64.72 &
  \cellcolor[HTML]{D1F5C7}-69.18 \\
\multirow{-4}{*}{CiteUL} &
  ItemKNN &
  \cellcolor[HTML]{D1F5C7}48.54 &
  \cellcolor[HTML]{D1F5C7}18.93 &
  \cellcolor[HTML]{D1F5C7}16.29 &
  \cellcolor[HTML]{D1F5C7}7.61 &
  \cellcolor[HTML]{D1F5C7}5.72 \\ \hline
 &
  HybridSVD &
  \cellcolor[HTML]{D1F5C7}87.46 &
  \cellcolor[HTML]{D1F5C7}48.37 &
  \cellcolor[HTML]{D1F5C7}16.12 &
  \cellcolor[HTML]{D1F5C7}12.45 &
  \cellcolor[HTML]{D1F5C7}5.59 \\
 &
  LCE &
  \cellcolor[HTML]{D1F5C7}81.24 &
  -13.06 &
  -0.83 &
  \cellcolor[HTML]{D1F5C7}21.67 &
  \cellcolor[HTML]{D1F5C7}27.42 \\
\multirow{-3}{*}{Amz G} &
  ItemKNN &
  \cellcolor[HTML]{D1F5C7}69.51 &
  \cellcolor[HTML]{D1F5C7}68.94 &
  \cellcolor[HTML]{D1F5C7}42.98 &
  \cellcolor[HTML]{D1F5C7}22.35 &
  \cellcolor[HTML]{D1F5C7}9.02 \\ \hline
 &
  HybridSVD &
  \cellcolor[HTML]{D1F5C7}48.82 &
  \cellcolor[HTML]{D1F5C7}44.79 &
  \cellcolor[HTML]{D1F5C7}36.23 &
  \cellcolor[HTML]{D1F5C7}11.65 &
  \cellcolor[HTML]{D1F5C7}{\ul 0.94} \\
 &
  LCE &
  \cellcolor[HTML]{D1F5C7}49.97 &
  \cellcolor[HTML]{D1F5C7}14.53 &
  \cellcolor[HTML]{D1F5C7}17.81 &
  \cellcolor[HTML]{D1F5C7}12.18 &
  \cellcolor[HTML]{EBF6E8}6.04 \\
\multirow{-3}{*}{Amz S} &
  ItemKNN &
  \cellcolor[HTML]{D1F5C7}86.05 &
  \cellcolor[HTML]{D1F5C7}67.62 &
  \cellcolor[HTML]{D1F5C7}44.49 &
  \cellcolor[HTML]{D1F5C7}20.17 &
  \cellcolor[HTML]{D1F5C7}8.46 \\ \hline
 &
  HybridSVD &
  \cellcolor[HTML]{D1F5C7}27.48 &
  \cellcolor[HTML]{D1F5C7}28.76 &
  \cellcolor[HTML]{EBF6E8}10.92 &
  \cellcolor[HTML]{EBF6E8}5.32 &
  \cellcolor[HTML]{D1F5C7}5.19 \\
 &
  LCE &
  \cellcolor[HTML]{D1F5C7}27.83 &
  -21.51 &
  -27.74 &
  -13.56 &
  \cellcolor[HTML]{EBF6E8}{\ul 7.87} \\
\multirow{-3}{*}{Amz IS} &
  ItemKNN &
  \cellcolor[HTML]{D1F5C7}83.12 &
  \cellcolor[HTML]{D1F5C7}73.59 &
  \cellcolor[HTML]{D1F5C7}54.35 &
  \cellcolor[HTML]{D1F5C7}22.44 &
  \cellcolor[HTML]{EBF6E8}3.58 \\ \hline
\end{tabular}%
}
\end{table}

On the Book Crossing and MTS Library datasets with non-textual features, our feature selection approach significantly outperforms the CFeCBF and random baselines. On average, our approach provides 12\% increase in metric for HybridSVD, 5\% for LCE, 20\% for LightFM, and 14\% for ItemKNN on the datasets with non-textual features, compared to the best-performing popular baseline. The strong performance of the popular feature selection baseline on datasets with non-textual features can be attributed to the sparsity of the feature matrix -- in these datasets, each item is typically described by only 2-3 features (out of 1,271 for MTS and 2,385 for BX). Selecting features with more interactions effectively includes more informative signals for the recommender model, resulting in a much higher metric value compared to random selection, where the feature matrix becomes even sparser.

On the datasets with textual features, our approach shows much better performance, improving recommendation quality by up to 95\%. The reason for such behavior could be the density of features, which is also indicated by the decent performance of the random feature selection approach, which, to our surprise, even outperformed the popularity-based approach.

 \subsubsection{Importance of Feature Categories}

\begin{figure}
\setlength{\abovecaptionskip}{5pt}
   \centering
   \includegraphics[width=\columnwidth]{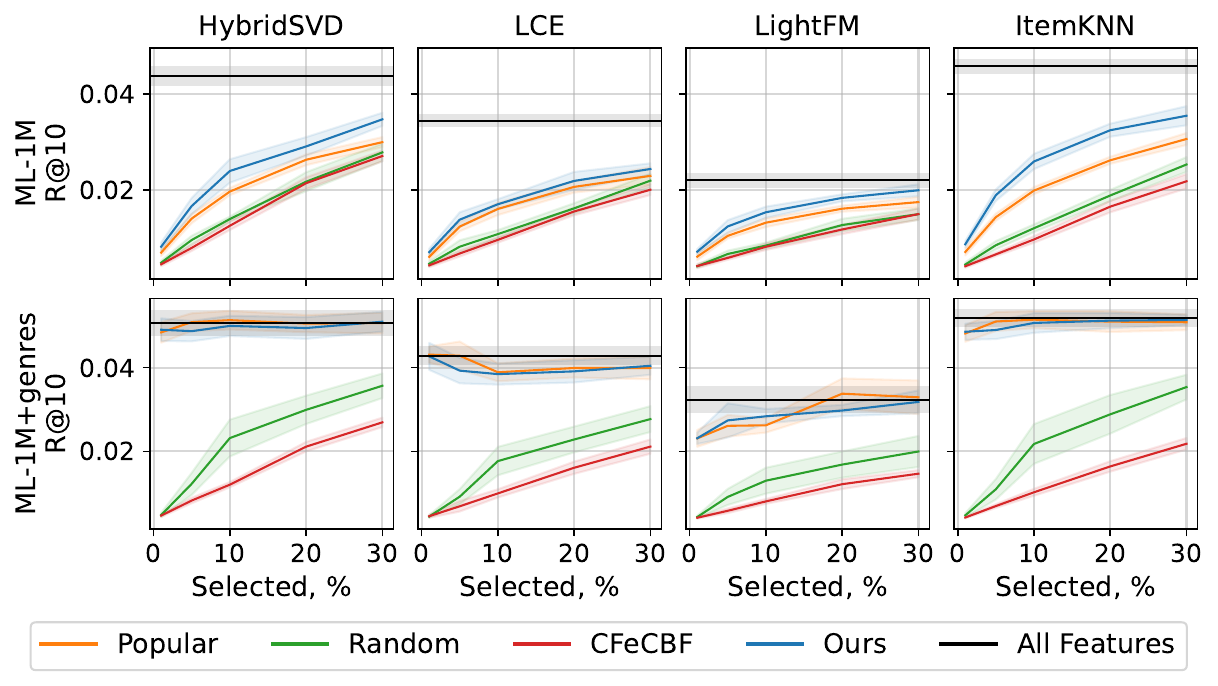}
   \Description{A figure with performance comparison of different feature selection algorithms.}
   \caption{Recall@10 comparison between excluding (top row) and including (bottom row) the Genres feature category for ML-1M.}
   \label{fig:ml-1m_recall}
 \end{figure}

We also aimed to assess how different feature categories influence recommendation quality. To achieve this, we tested our approach using the ML-1M dataset, due to its nonuniform feature distribution. The dataset includes 2,774 actor, 731 director, and 160 writer features, along with a genre category containing 21 features. We conducted experiments both with and without the genre category. When genres were included, the HybridSVD, LCE, and ItemKNN models achieved the same quality as the full-feature approach, but using only 1\% of the features (Fig.~\ref{fig:ml-1m_recall}). Of the 36 features selected (representing 1\% of the total), our approach chose 13 of the 21 genres, indicating that the genres category strongly drives the recommendations. The popularity-based method selected 16 genre features and demonstrated the same quality as our approach. On the other hand, random and CFeCBF feature selection baselines did not include any features from genres category, resulting in poor performance of the model at low percentage of selected features.

We attribute this behavior to the dominance of the genre category, which is the most frequently occurring feature type. On average, genres appear in $303.5$ movies, compared to $3.1$ for actors, $2.5$ for directors, and $1.5$ for writers. Consequently, the genres contain significantly more (up to $100$ times) information than the other categories combined, causing the recommender to overfit to this signal. This was further validated by the experiment without the genres category, where the quality of models using 1\% of features dropped significantly compared to those using all available features.

\begin{figure}
 \setlength{\abovecaptionskip}{4pt}
   \centering
   \includegraphics[width=\columnwidth]{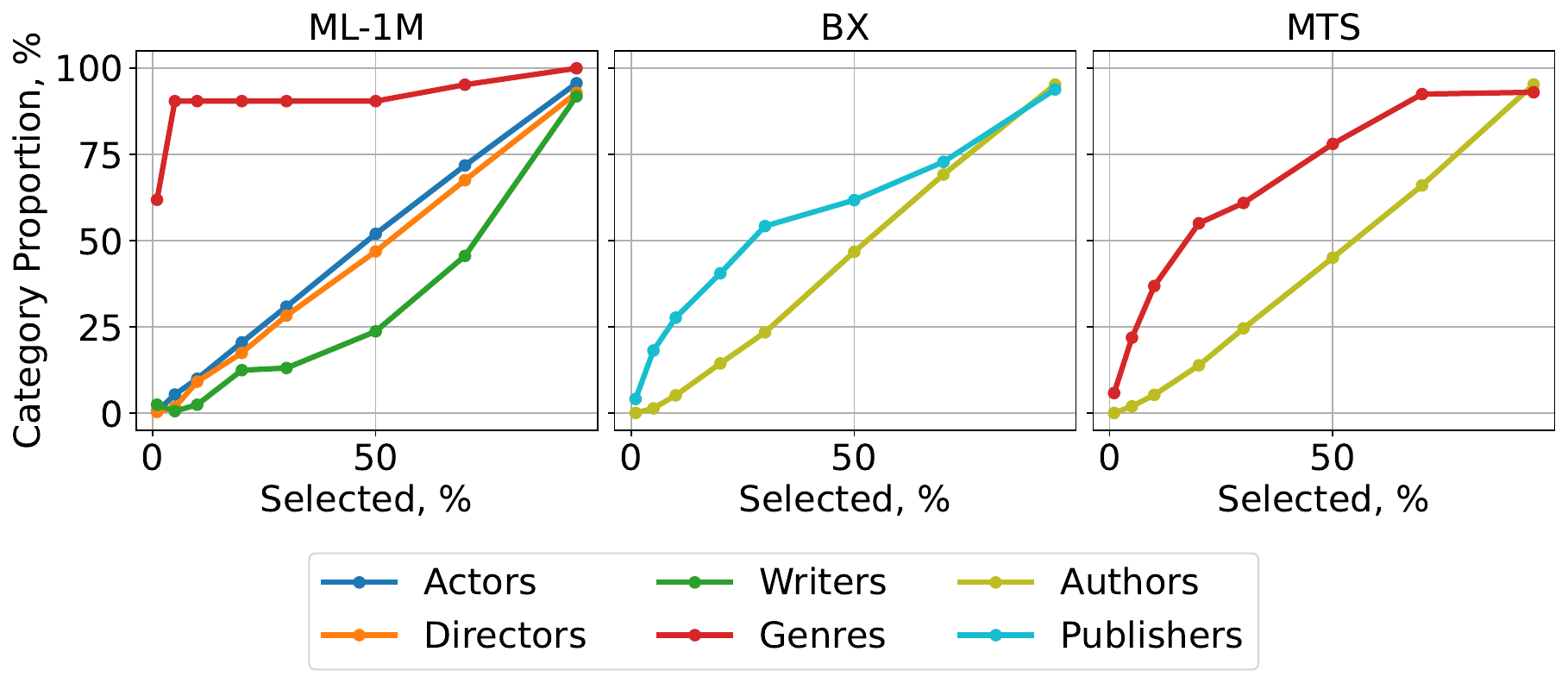}
   \Description{A figure with performance comparison of different feature selection algorithms.}
   \caption{The proportion of selected features inside each category for the datasets with non-textual features.}
   \label{fig:ml-1m_categories}
 \end{figure}

We also examined the selection of features from each category. Since only datasets with non-textual features have distinct categories (textual datasets contain tokens that are hard to cluster), we plotted the proportion of selected features from each category across different fractions of selected features (Figure~\ref{fig:ml-1m_categories}). This plot serves as an indicator of the importance of feature categories. For ML-1M, the \textit{genres} category was almost entirely selected, indicating its high importance. The \textit{actors} and \textit{directors} categories showed a linear increase in selection as the overall number of selected features grew, while \textit{writers} category had the lowest selection rate. This suggests that the \textit{genres} is the most critical category, while the \textit{writers} is the least important. A similar pattern was observed for the MTS and BX datasets, where features from the \textit{authors} category were selected less frequently than those from other categories. Our approach can, therefore, be used to determine importance of different feature categories based on how actively features from categories are selected.

\subsubsection{Hyperparameter Analysis}

 \begin{figure}[b]
 \setlength{\abovecaptionskip}{4pt}
   \centering
   \includegraphics[width=\columnwidth]{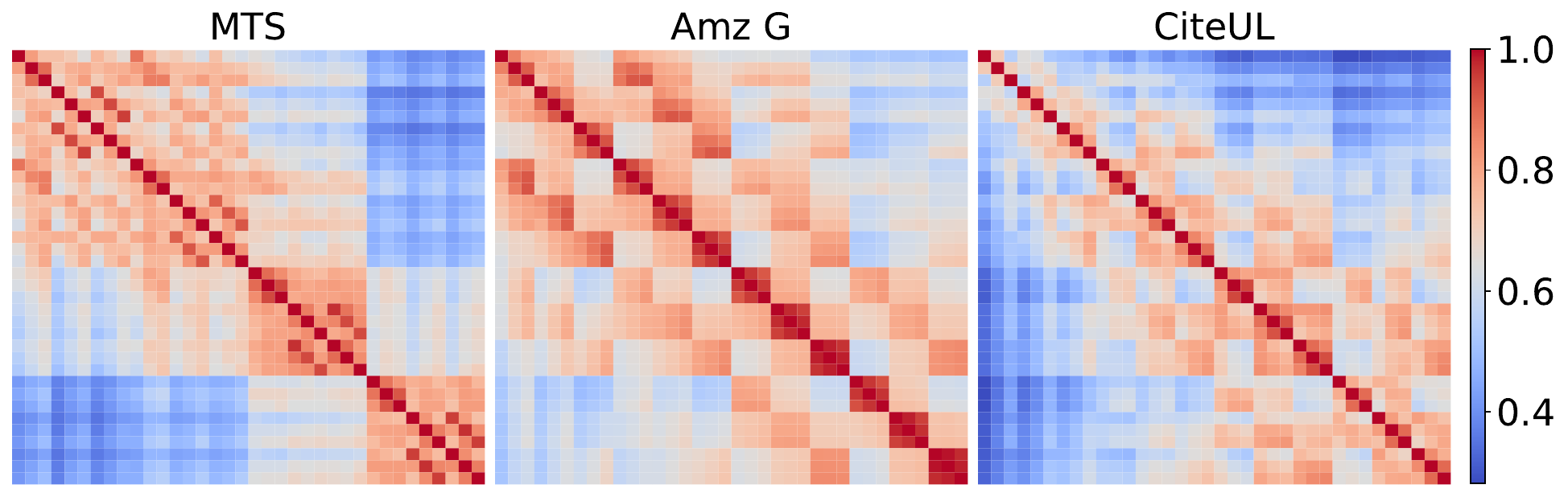}
   \Description{A figure with performance comparison of different feature selection algorithms.}
   \caption{Jaccard similarity coefficient between selected feature subsets for different hyperparameter configurations. 10\% of features are selected for each dataset.}
   \label{fig:jaccard}
 \end{figure}

We used the following ranges for the hyperparameters based on preliminary testing: $\alpha \in [0.2, 0.5, 0.8]$, $p \in [-1.0, -0.5, 0.0, 0.5]$, $k \in [100, 200, 400]$. To study the sensitivity of these hyperparameters, we selected the Amazon Video Games, CiteULike, and MTS Library datasets to cover both textual and non-textual features. We chose 10\% of feature indices using each configuration of hyperparameters and measured the Jaccard similarity coefficient~\cite{han2011jaccard} between each pair of feature index sets, visualizing the results in heatmaps (Fig.~\ref{fig:jaccard}).

The datasets displayed distinct behaviors. For instance, MTS dataset formed three clusters, squares of red areas on main diagonal. These clusters correspond to different values of hyperparameter $p$: the upper left square area refers to $p\in [-0.5, -1.0]$, the middle square to $p=0.0$, and the lower right large square to $p=0.5$. This means that the popularity hyperparameter $p$ has the most significant impact on feature selection results. AMZ G dataset is primarily influenced by the $p$ and $k$ hyperparameters. The heatmap illustrates numerous red areas in the form of 3$\times$3 squares, each corresponding to specific combinations of $p$ and $k$. This indicates that the impact of the $\alpha$ hyperparameter is relatively minimal when $p$ and $k$ have higher values, as evidenced by the patterns in the lower right corner of the plot. The CiteULike dataset exhibited minimal stratification in terms of feature selection hyperparameters. There are some clusters associated with nonnegative values of $p$, but they are not so distinct as those observed in the other datasets.

From these findings, we conclude that the item popularity hyperparameter $p$ has the most significant influence on the set of selected features; the hyperparameter $k$, the embedding size, has a moderate impact; and the $\alpha$ hyperparameter, which is responsible for the contribution of the collaborative signal, has the least impact.

\subsubsection{Ablation Study}

The first part of ablation study focuses on the enrichment of features with collaborative information -- first step of our feature selection algorithm. We investigated whether incorporating a collaborative signal (controlled by $\alpha$) enhances recommendation quality or not. The results, presented in Table~\ref{tab:ablation_colab}, are statistically significant. We tested this hypothesis using the HybridSVD and ItemKNN models due to their superior efficiency. Our findings indicate that addition of user behavioral signals positively impacts recommendation quality compared to the scenario where $\alpha=0$, which relies solely on content features without any collaborative signal for feature selection. Even a small amount of collaborative information consistently improves the performance metrics across both models, demonstrating the importance of integrating user behavior data into the feature selection process.

\begin{table}[]
  \setlength{\abovecaptionskip}{3pt}
  \caption{Ablation study: effects of collaborative data in recommendation quality (R@10) for different \% selected features.}
  \resizebox{\columnwidth}{!}{%
    \begin{tabular}{l l rrrr rrrr}
      \toprule
      \multirow{2}{*}{Dataset} 
        & \% Selected 
        & \multicolumn{4}{c}{HybridSVD} 
        & \multicolumn{4}{c}{ItemKNN} \\
      \cmidrule(l{1em}r{1em}){3-6}
      \cmidrule(l{1em}r{1em}){7-10}
        & $\rightarrow$ 
        & 1\% & 5\% & 10\% & 30\% 
        & 1\% & 5\% & 10\% & 30\% \\
      \hline
      \multirow{3}{*}{CiteUL} 
        & $\alpha \neq 0$  & 0.105 & 0.154 & 0.169 & 0.191 & 0.107 & 0.145 & 0.168 & 0.185 \\
        & $\alpha=0$       & 0.086 & 0.133 & 0.152 & 0.183 & 0.080 & 0.124 & 0.147 & 0.178 \\
        & \textit{Effect}  & \textit{-18\%} & \textit{-14\%} & \textit{-10\%} & \textit{-4.2\%}
                          & \textit{-25\%} & \textit{-15\%} & \textit{-13\%} & \textit{-3.8\%} \\
      \midrule
      \multirow{3}{*}{MTS}   
        & $\alpha \neq 0$  & 0.113 & 0.187 & 0.234 & 0.296 & 0.113 & 0.177 & 0.203 & 0.235 \\
        & $\alpha=0$       & 0.104 & 0.174 & 0.220 & 0.290 & 0.104 & 0.165 & 0.189 & 0.222 \\
        & \textit{Effect}  & \textit{-8.0\%} & \textit{-7.0\%} & \textit{-6.0\%} & \textit{-2.0\%}
                          & \textit{-8.0\%} & \textit{-6.8\%} & \textit{-6.9\%} & \textit{-5.5\%} \\
      \bottomrule
    \end{tabular}%
  }
  \label{tab:ablation_colab}
\end{table}

This part of ablation study examines the feature ranking and the applicability of MaxVol algorithm for selecting optimal feature sets. We compare MaxVol against two alternatives: random selection (baseline) and norm selection, where feature embeddings derived at the end of Section~\ref{sec:feature_selection} were sorted by their $L_2$ norm (norm selection). Results in Table~\ref{tab:abl_maxvol} show that MaxVol clearly outperforms both methods when only 1\% of features are selected. As the percentage of selected features increases, MaxVol and norm selection continue to outperform random selection. However, at higher percentiles, the performance gap between MaxVol and norm selection becomes statistically insignificant. Due to space constraints, results for higher percentiles are omitted. Overall, MaxVol proves most effective in extreme selection settings, making it well-suited for scenarios that demand compact yet representative feature subsets.

\subsubsection{Recommender Model Runtime Analysis}

 \begin{figure}[tbh]
\setlength{\abovecaptionskip}{3pt}
   \centering
   \includegraphics[width=0.85\columnwidth]{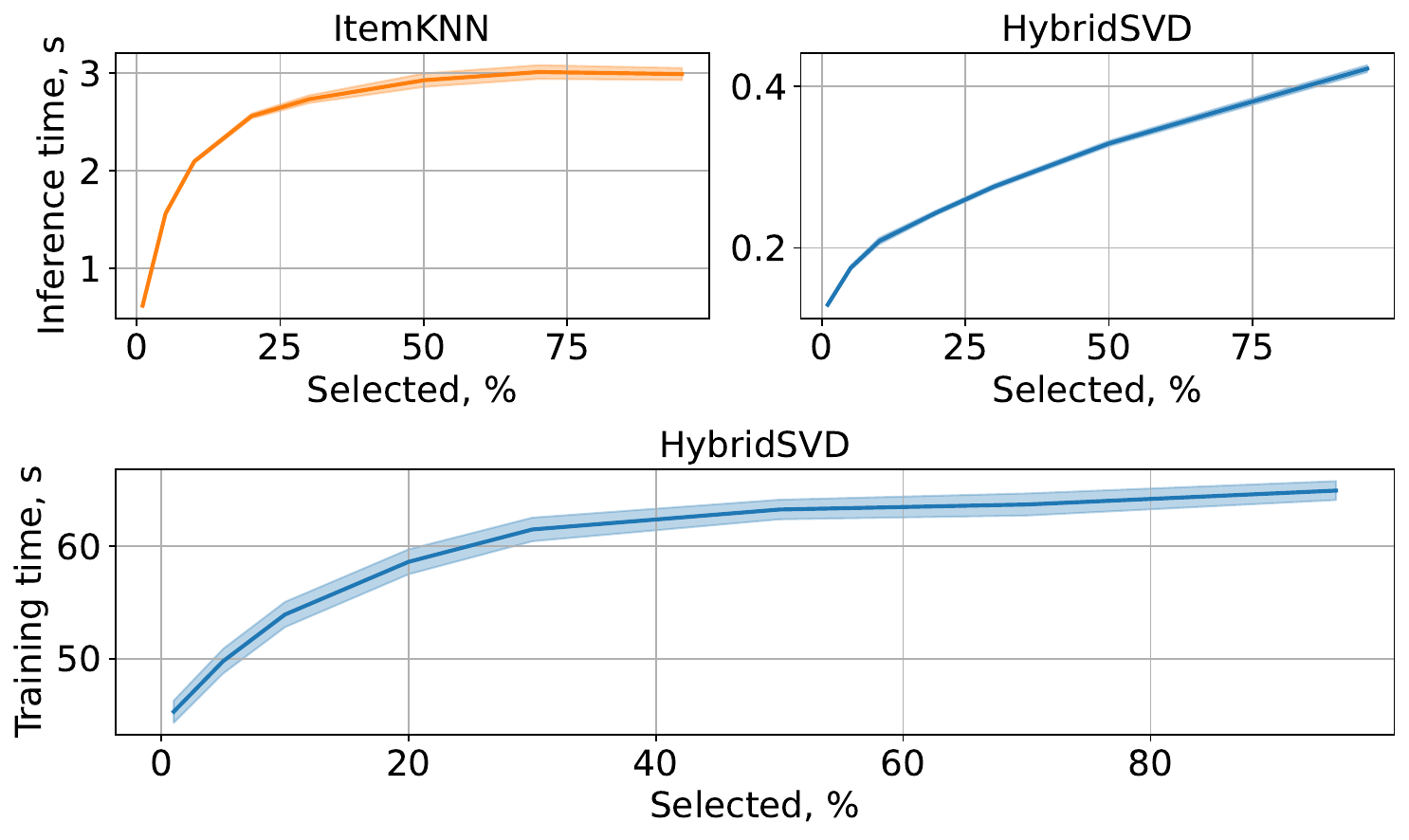}
   \Description{Training and inference time.}
   \caption{Inference time (top row) and training time (bottom row) for different models on the Amazon Video Games dataset for different percentiles of selected features. For the ItemKNN model all the computations are performed only at the inference step.}
   \label{fig:time_ablation}
 \end{figure}

Figure~\ref{fig:time_ablation} illustrates the average training and inference times for the ItemKNN and HybridSVD models as a function of the number of features used. Notably, when using just 5\% of features, the ItemKNN model has the same recommendation quality as using all 100\% of features (see Fig.~\ref{fig:results_recall}), while achieving 2$\times$ speedup in terms of computational efficiency. When utilizing only 1\% of the features, the ItemKNN model demonstrates a 5$\times$ speedup in inference time, while the HybridSVD model shows a 3$\times$ speedup in inference time and a 1.4$\times$ speedup in training time. These significant improvements in computational performance can be attributed solely to the reduction of the feature space. Our approach not only maintains the recommendation quality, but also achieves substantial gains in computational efficiency, making the recommender systems more scalable and responsive.

\begin{table}[tbh]
\setlength{\abovecaptionskip}{3pt}
\caption{Ablation study: effects of feature selection methods on recommendation quality (R@10) for 1\% selected features.}
\resizebox{\columnwidth}{!}{%
\begin{tabular}{l|l|rrr|rrr}
\hline
\multirow{2}{*}{Dataset} & Method          & \multicolumn{3}{c|}{\textbf{HybridSVD}} & \multicolumn{3}{c}{\textbf{ItemKNN}} \\
 &
  \multicolumn{1}{r|}{$\rightarrow$} &
  \multicolumn{1}{c}{MaxVol} &
  \multicolumn{1}{c}{Random} &
  \multicolumn{1}{c|}{Norm} &
  \multicolumn{1}{c}{MaxVol} &
  \multicolumn{1}{c}{Random} &
  \multicolumn{1}{c}{Norm} \\ \midrule
\multirow{2}{*}{BX}      & R@10            & 0.041    & 0.002    & 0.037    & 0.040   & 0.002   & 0.035   \\
                         & \textit{Effect} & --       & -\textit{94\%}    & -\textit{9.8\%}   & --      & -\textit{95\%}   & -\textit{13\%}   \\ \midrule
\multirow{2}{*}{MTS}     & R@10            & 0.113    & 0.005    & 0.051    & 0.113   & 0.005   & 0.056   \\
                         & \textit{Effect} & --       & \textit{-95\%}    & \textit{-55\%}    & --      & \textit{-95\%}   & \textit{-50\%}   \\ \hline
\end{tabular}%
}
\label{tab:abl_maxvol}
\end{table}
\section{Related Work}
\label{sec:relatedwork}
Feature handling in recommender systems includes feature selection~\cite{fs_survey, beraha2019feature, yasuda2022sequential}, feature weighting~\cite{featureweighting}, and dimensionality reduction~\cite{dimreduction}. Feature selection methods are typically categorized as filter, embedded, or wrapper approaches~\cite{fs_survey, microsoft1, microsoft2}.

\emph{Wrapper methods}, which our approach follows, evaluate subsets of features by retraining the model. \citet{FSforFMRecsys} proposed to use the weights of a trained FM model to identify the most important features, with higher weights indicating greater importance. Their method was applied to a regression task focused on predicting session time, which is distinct from a ranking task.
\citet{Deldjoo_cfecbf} addressed the cold-start problem in movie recommendations. They propose a feature weighting based on the approximation of the item similarity matrix constructed from a pre-trained collaborative model with an item similarity matrix acquired from a content-based model. While this approach effectively learns feature weights using collaborative signal, the authors did not investigate its potential for feature selection. In our work, we use these learned weights as feature importances, sorting the features accordingly, to compare the performance with our feature selection approach.
A sophisticated approach was developed by \citet{Nembrini_CQFS} as a development of ideas from the feature weighting approach~\cite{Deldjoo_cfecbf}. Authors framed feature selection as a QUBO problem solved on quantum hardware, later approached classically by \citet{Nikitin_TTOPT}, though with limited replication of results. In some cases, the approach, contrary to the original work, suggested picking all available features, disregarding the restriction on the number of selected features due to the soft (not hard) nature of the constraint in the QUBO problem.

Our method differs by directly incorporating collaborative signals into content feature embeddings and applying MaxVol for selection. It enforces a hard constraint on the number of selected features, requires no specialized hardware, and is well-suited for cold-start scenarios in simple, interpretable linear models, as feature selection in recommender systems remains an underexplored area. We also highlight that basic baselines, such as random or popularity-based selection, can be surprisingly competitive yet are often overlooked. By addressing this gap, our work aims to provide a robust foundation for advancing research and practice in feature selection for recommender systems.

\section{Conclusion}
\label{sec:conclusion}

We proposed a novel feature selection algorithm that addresses the cold-start problem in recommender systems by leveraging collaborative signals to rank feature importance. By embedding behavioral correlations into the feature space and applying a structured ranking mechanism, our method effectively identifies the most relevant features, achieving strong recommendation performance with a significantly reduced feature set. Experiments across seven diverse datasets demonstrate that our approach consistently outperforms existing feature selection techniques while also improving computational efficiency. The method is model-agnostic, scalable, and practical for hybrid recommendation scenarios. 

\bibliographystyle{ACM-Reference-Format}
\balance
\bibliography{bibliography}

\end{document}